\documentclass[pra,superscriptaddress,onecolumn,floatfix]{revtex4}
\usepackage{amsbsy,amsmath,amsfonts,latexsym}
\usepackage{bm,color,subfigure,graphics}
\usepackage{srcltx,epsfig,float,wrapfig}
\def\bra#1{\mathinner{\langle{#1}|}}
\def\ket#1{\mathinner{|{#1}\rangle}}

\begin{document}
\title{Spontaneous emission in the presence of incident fields}  
\author{Peter G. Brooke}
\email{pgb@ics.mq.edu.au}
\affiliation{Centre for Quantum Computer Technology and Department of Physics,
Macquarie University, Sydney, New South Wales 2109, Australia}
\date{\today}
\begin{abstract}
We characterise the spontaneous emission time and direction from small numbers
of dipole-coupled two-level atoms (2LAs) in the presence of incident fields.
We show how to use adiabatic passage to admit population transfer between
states in the one-quantum subspace for two and three 2LAs.  Our method is a  
multi-atom generalisation of stimulated-Raman-adiabatic-passage (STIRAP) for
a single multi-level atom.  We use numerical results to justify an
ansatz that enables us to give analytical expressions for the directional
emission which depends on the incident fields.  Our results
admit a characterisation of the efficacy of population transfer in small
numbers of dipole-coupled 2LAs, and are applicable to proof-of-principle
experiments involving dipole-coupled 2LAs. 
\end{abstract}
\maketitle
%%%%%%%%%%%%%%%%%%%%%%%%%%%%%%%%%%%%%%%%%%%%%%%%%%%%%%%%%%%%%%%%%%%%%%%%%%%%%
\section{Introduction}
%%%%%%%%%%%%%%%%%%%%%%%%%%%%%%%%%%%%%%%%%%%%%%%%%%%%%%%%%%%%%%%%%%%%%%%%%%%%
Atomic systems have proven to be valuable for experimental 
demonstrations of entanglement and violations of Bell
inequalities~\cite{Rowe01}, quantum information
protocols~\cite{Mon95,Ste97,Wei02}, cooperative effects such as
superradiance~\cite{DeV96}, and quantum beats~\cite{Vit02}.
A pair of electric dipole-coupled 
two-level atoms (2LAs) has been especially well studied~\cite{Ficek02}, as
this is the simplest atomic system for demonstrating 
these effects, and three 2LAs is 
the smallest number of atoms that encode a decoherence-free qubit (two-level
system)~\cite{Kem01}. Common to all demonstrations of 
quantum beats, cooperative effects, entanglement, and quantum processing tasks
is the need for external control.  In most physical systems, external control
requires incident fields, which can alter the characteristics of the system.
Quantifying the effect of such fields on the emission properties of small
collections of 2LAs is one purpose of this paper. 

We focus on systems in which the atomic coupling is described by a
dipole-dipole interaction.  We assume the 2LAs are closely separated, and so
permit only global control.  Previous work has quantified global preparation,
manipulation, and readout of collective states~\cite{Bei99,Brooke07}. 
Here, we show how to transfer population between entangled states in both two
and three 2LAs using resonant counter-intuitive pulses similar to 
stimulated Raman adiabatic passage (STIRAP)~\cite{Berg98}.  There are two
reasons why it is surprising that this technique is applicable.  First, there
is no null eigenstate in the field Hamiltonian---normally required for
STIRAP---and second, the 2LAs are spatially separated.   Our results exploit
the energy-level splitting due to the dipole-dipole interaction,  allowing
certain (fast-oscillating) terms to be neglected.  This enables an
eigenstate of the field Hamiltonian to be approximated by a dark-state.  The
approximate dark state is between two maximally-entangled states, so we call
our method entangled-STIRAP.   For two 2LAs, entangled-STIRAP is possible at
large separations, and so we hope is amenable with present technology. 
  
We then examine the spontaneous emission characteristics in the presence of
global control-fields incident on two and three 2LAs.  Quantifying the emission
characteristics serves two purposes.  First, the 
decay timescale allows for the detrimental effect of the control-fields on the
emission timescale to be characterised.  Second, the emission direction
provides a means with which to quantify the efficacy of the transfer.  This
gives a practical way to measure the fidelity of the population transfer.  

The paper is structured as follows.  In Sec.~\ref{sec:physsys}, we describe
the physical system, and a particular unravelling of the Lindblad master
equation.  Then, in Sec.~\ref{sec:tat}, we focus on two 2LAs.  In
Sec.~\ref{sec:manip2a}, we propose two methods for coherent rotation
in the one-excitation subspace: collective two-photon Raman transitions
and entangled-STIRAP.  In Sec.~\ref{sec:emit2a}, we use numerical results
to justify an ansatz that enables us to find an analytical expression
for the directional emission matrix element in the presence of an incident
field.  We then turn our attention to three 2LAs.  In
Sec.~\ref{sec:manip3a}, we show how to use entangled-STIRAP to rotate
information encoded in the slowest decaying excited-states in three 2LAs.
Then, in Sec.~\ref{sec:emit3a}, we give an indication of the effect of the
incident fields on the emission pattern and time, and we show numerically how
the emission pattern is affected by the incident fields.  
%%%%%%%%%%%%%%%%%%%%%%%%%%%%%%%%%%%%%%%%%%%%%%%%%%%%%%%%%%%%%%%%%%%%%%%%%%%%%
\section{Physical system}
\label{sec:physsys}
%%%%%%%%%%%%%%%%%%%%%%%%%%%%%%%%%%%%%%%%%%%%%%%%%%%%%%%%%%%%%%%%%%%%%%%%%%%%
Our analysis is in accordance with standard quantum optical methods for
electric dipole-coupled two-level atoms (2LAs): a quantum master equation
employing the 
rotating-wave and Born-Markov approximations.   We assume that the spatial 
extent of the 2LA is much less than the resonant wavelength $\lambda_0 = 2 \pi
/k_0$, and so the 2LAs are effectively point dipoles with resonant
frequency $\omega_0$.  We describe the driving
field as a classical bichromatic field that drives all 2LAs simultaneously,
but due to the small interatomic separations cannot drive each 2LA
individually.  The raising operator for atom 
$i$ is $\hat{\sigma}_{i+}=|1\rangle_{i }\langle 0| =
\hat{\sigma}_{i-}^\dagger$ 
and $\hat{\sigma}_{iz}=[\hat{\sigma}_{i+}, \hat{\sigma}_{i-}]$.
The transition-matrix element of the atom is given by $\boldsymbol{d} = \,
_i\!\bra{0}\hat{\boldsymbol{d}}_i \ket{1}_i$  with the dipole operator 
$\hat{\boldsymbol{d}}_i$ for atom $i$.  The dipoles of the atoms are
identically oriented, with $\vec{d} \cdot \vec{r}_{ij} = \cos \alpha$ for
$\vec{d}$ the unit vector in direction $\boldsymbol{d}$, and $\vec{r}_{ij} =
\boldsymbol{r}_{ij} /  r_{ij} $ the unit vector in the direction of the
separation 
vector $\boldsymbol{r}_{ij}$ between atoms $i$ and $j$.
Since all the atoms 
are identical, the matrix elements of their dipole operators are equal:
$\boldsymbol{d} = \boldsymbol{d}_i \forall i$.  

The free evolution of $N$ dipole-coupled atoms, including all
non-nearest-neighbour interactions, without an incident field is $(\hbar = 
1)$~\cite{Car93}     
\begin{align}
\label{eq:hnh}
\hat{H}_{\text{S}} = \frac{\omega_0}{2} \sum_{i=1}^N \hat{\sigma}_{iz} +
  \sum_{i \ne 
  j=1}^N \Xi_{ij} \hat{\sigma}_{i+} \hat{\sigma}_{j-} - \text{i}
  \frac{\gamma}{2} \sum_{i = 1}^N  \hat{\sigma}_{i+} \hat{\sigma}_{i-}, 
\end{align}
with 
\begin{align}
\Xi_{ij} \equiv -\frac{3 \gamma}{4}
\frac{\text{e}^{\text{i}\xi_{ij}}}{\xi_{ij}^{3}} \left[ \xi_{ij}^{ 2} \sin^{2}
  \alpha  - 
  \left( 1 - \text{i}\xi_{ij} \right) \left( 1-3\cos^{2} \alpha \right)
  \right], 
\end{align}
for $i,j = 1, \ldots, N$, $\xi_{ij} \equiv k_0 r _{ij}$, and $\gamma$ the
single-atom decay rate.  By symmetry, $\Xi_{ij} = \Xi_{ji}$, and we define
$\Delta_{ij} \equiv \text{Re}\{\Xi_{ij}\}$ and $\gamma_{ij} 
\equiv -2 \, \text{Im} \{\Xi_{ij}\}$.  The coefficients $\Delta_{ij}$ and
$\gamma_{ij}$ correspond to the dipole-dipole interaction and the
actual process of photon emission respectively.  The $i^\text{th}$ atom is
located at $\boldsymbol{r}_i = (i-1) s \lambda_0 \vec{z}$, for
$\vec{z}$ the unit vector along the interatomic axis, and $s$ the
interatomic separation in units of emission wavelength. 

The laser has a bichromatic electric field $\boldsymbol{E}(\boldsymbol{r}) =
\boldsymbol{E}_\mu(\boldsymbol{r}) + \boldsymbol{E}_\nu(\boldsymbol{r})$, with
$\boldsymbol{E}_\mu(\boldsymbol{r})$ and $\boldsymbol{E}_\nu(\boldsymbol{r})$
the electric field amplitudes. $\boldsymbol{E}(\boldsymbol{r})$ interacts with
the atoms via a dipole coupling and so we introduce the Rabi
frequencies $\mathcal{E}_{\mu,i} = \boldsymbol{d} \cdot \boldsymbol{E}_\mu
\text{e}^{-\text{i}   \boldsymbol{k}_\mu \cdot \boldsymbol{r}_{i}}$ and $
\mathcal{E}_{\nu,i} = 
\boldsymbol{d} \cdot \boldsymbol{E}_\nu \text{e}^{-\text{i} \boldsymbol{k}_\nu
  \cdot \boldsymbol{r}_{i}} $ for wavevectors $\boldsymbol{k}_\mu$ and
$\boldsymbol{k}_\nu$. Within the rotating-wave approximation, the interaction
Hamiltonian is   
\begin{align}
\hat{H}_{\text{I}} =  \sum^{N}_{i=1} \mathcal{E}_{i} \hat{\sigma}_{i-}
+ \text{H.c.},
\end{align}
for which $\mathcal{E}_{i}$ are described by a time-dependent bichromatic
external field:  
\begin{align}
 \mathcal{E}_{i}  = \mathcal{E}_{\mu,i} \text{e}^{\text{i} \omega_\mu t } + 
\mathcal{E}_{\nu,i} \text{e}^{\text{i} \omega_\nu t }.   
\end{align}
Due to the different positions of the atoms, the field $\mathcal{E}_{i}$
differs for distinct atoms.  The total effective Hamiltonian for the no-jump
evolution is $\hat{H}_{\text{eff}} = \hat{H}_{\text{S}} +
\hat{H}_{\text{I}}$.   

The jump operators are written~\cite{Car00}
\begin{align}
\hat{S}(\theta, \phi) = \sqrt{\gamma D(\theta,\phi) d \Omega} \sum_{j=1}^N
\text{e}^{-\text{i} k_0 \vec{R}(\theta, \phi) \cdot \boldsymbol{r}_j}
\hat{\sigma}_{j-}, 
\end{align}
which apply when a photon is detected in the far field within the element of
solid angle $d\Omega$ in direction $\vec{R}(\theta, \phi)$. The dipole
radiation pattern is
\begin{align}
D(\theta, \phi) = \frac{3}{8 \pi} \{ 1 - [\vec{d} \cdot \vec{R}(\theta,
  \phi)]^2 \},
\end{align}
which is the directional emission from an isolated atom.  So, the master
equation is written
\begin{align}
\dot{\rho} = -\text{i}(\hat{H}_{\text{eff}} \rho - \rho
\hat{H}^\dagger_{\text{eff}}) + \int \hat{S}(\theta, \phi) \rho \hat{S}^\dagger
(\theta, \phi),
\end{align}
which corresponds with the standard Lehmberg master equation for $N$
atoms~\cite{Bela69,Lehm70i,Arg70}.  Writing the master equation in this way is
useful for analysing spontaneous emission in the presence of incident fields. 

When studying directional emission properties we only consider
$\alpha = \pi /2$.  So, the jump operators can be
written 
\begin{align}
\hat{S}(\theta) = \sqrt{\gamma D(\theta) \sin \theta d \theta} \sum_{j=1}^N
\text{e}^{-\text{i} 2 \pi (j-1) s \cos \theta} \hat{\sigma}_{j-},
\end{align}
where we have exploited the symmetry of the atomic system and integrated
over azimuthal angle.  In this instance,  the dipole radiation pattern 
is written
\begin{align}
D(\theta) = \frac{3}{4} - \frac{3}{8} \sin^2 \theta.
\end{align}
%%%%%%%%%%%%%%%%%%%%%%%%%%%%%%%%%%%%%%%%%%%%%%%%%%%%%%%%%%%%%%%%%%%%%%%%%%%%%
\section{Two atoms}
\label{sec:tat}
%%%%%%%%%%%%%%%%%%%%%%%%%%%%%%%%%%%%%%%%%%%%%%%%%%%%%%%%%%%%%%%%%%%%%%%%%%%%
We define the basis of two atoms as $\ket{\sf a}=\ket{00}$, corresponding
with both 2LAs in the ground state, and $|{\sf d}\rangle=\ket{11}$ plus the
symmetric state $|{\sf c}\rangle=(1/ \sqrt{2})( \ket{10} + \ket{01})$
and antisymmetric state $|{\sf b}\rangle= (1/ \sqrt{2})( \ket{10} -
\ket{01})$. 
The corresponding energies are $0, 2\omega_0, \omega_0+\Delta$ and
$\omega_0-\Delta$, respectively. The frequency difference between levels
$|{\sf a}\rangle$ and $|{\sf b}\rangle$ is given by $\omega_{\sf ab}$, and
similar notation is applied to other transitions.  For two atoms, $\Delta
\equiv \text{Re}\, \{\Xi_{12} \}$, $\Gamma \equiv -2 \, \text{Im}
\, \{ \Xi_{12} \}$, and the decay rates of the states $\ket{\sf b}$ and
$\ket{\sf c}$ are $\gamma - \Gamma$ 
and $\gamma + \Gamma$ respectively.  Note that the decay rate for the {\sf b-a}
transition approaches zero in the limit $\xi \to 0$.  The atoms are
arranged according to $\boldsymbol{r}_{1}= 0$ and $\boldsymbol{r}_{2}=
\boldsymbol{r}$.   
%%%%%%%%%%%%%%%%%%%%%%%%%%%%%%%%%%%%%%%%%%%%%%%%%%%%%%%%%%%%%%%%%%%%%%%%%%%%%
\subsection{Manipulation}
\label{sec:manip2a}
%%%%%%%%%%%%%%%%%%%%%%%%%%%%%%%%%%%%%%%%%%%%%%%%%%%%%%%%%%%%%%%%%%%%%%%%%%%%
\begin{figure}[t]
\begin{center}
\subfigure[]{\label{fig:ramantwo}
\includegraphics[width=6cm]{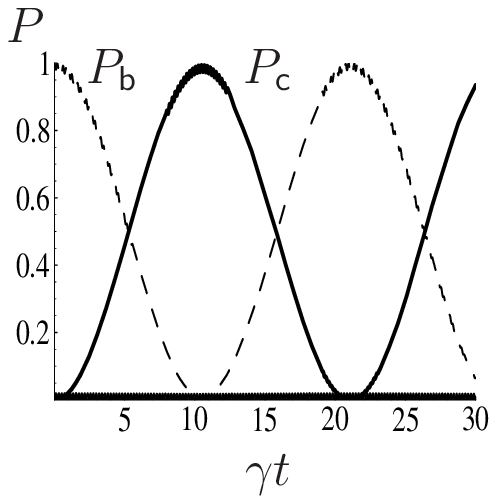}}
\subfigure[]{\label{fig:stiraptwo}
\includegraphics[width=6cm]{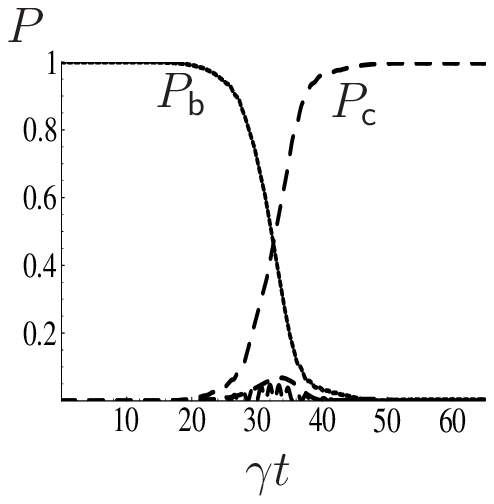}}
\end{center}
\caption{ Populations of the levels {\sf a,b,c,d} when
  $\hat{H}_{\text{eff}}$ from Eq.~\eqref{eq:hamtot} is applied to two atoms
 initially in $\ket{\sf b}$ with (a) $\mathcal{E}_\mu = \mathcal{E}_\nu =
  3\gamma$, $\omega_\delta  = 30 
  \gamma$, $\alpha = 0$, and $\xi_{12} = 1/5$, and (b)
  $\mathcal{E}_\mu(t) = 
  (3/4) \gamma \exp[- (1/5) (t - t_\mu)^2/T^2]$,
  $\mathcal{E}_\nu(t) = 
  (3/4) \gamma \exp[- (1/5) (t - t_\nu)^2/T^2]$, for $T =
  5.5 \gamma^{-1}$, $t_\mu = 4.5 T$, $t_\nu = t_\mu +
  2.75T$, $\alpha = 0$, $\omega_\delta  = 0$, and $\xi_{12} = 1.25$.}         
\end{figure}
Bell state preparation of two dipole-coupled 2LAs has been studied both 
experimentally~\cite{Tuc98,Roos04} and
theoretically~\cite{Bei99,Ficek03,Ficek04}. In particular, in
Ref.~\cite{Bei99} preparation of Bell states is performed by coherent
manipulation via an incident laser field.  We use the same preparation method
here, and we ignore decay for the rest of Sec.~\ref{sec:manip2a}.   For
preparation, 
only a single laser field is required.  So, in the interaction picture with
respect to the Hermitian part of $\hat{H}_{\text{S}}$, the Hamiltonian
$\hat{H}_{\text{eff}}$ becomes  
\begin{align}
\hat{H}_{\text{eff}} =& \frac{1}{2\sqrt{2}} \Big[(\mathcal{E}_{\mu,1} +
  \mathcal{E}_{\mu,2}) \Big( \text{e}^{-\text{i} \Delta t } \ket{ \sf a}
  \bra{\sf c} + \text{e}^{\text{i} \Delta t } \ket{ \sf c} \bra{\sf d} \Big)
- (\mathcal{E}_{\mu,1} -  \mathcal{E}_{\mu,2}) \Big( \text{e}^{\text{i}
  \Delta t } \ket{ \sf a} 
  \bra{\sf b} -  \text{e}^{-\text{i} \Delta t } \ket{ \sf b} \bra{\sf d} \Big)
  \Big] \text{e}^{\text{i}(\omega_\mu - \omega_0) t} + \text{H.c.},
\end{align}  
where $\mathcal{E}_{\mu,1}$ and $\mathcal{E}_{\mu,2}$ are the Rabi frequencies
at atoms one and two respectively. To 
prepare $\ket{\sf c}$, the incident fields are set to satisfy
$\mathcal{E}_{\mu,1} = \mathcal{E}_{\mu,2} = \mathcal{E}_{\mu}$ and the laser 
frequency tuned to $\omega_\mu = \omega_{\sf ab} = \omega_0 + \Delta$, which
gives   
\begin{align}
\hat{H}_{\text{eff}} = \frac{ \mathcal{E}_{\mu}}{\sqrt{2}} \left( 
\ket{\sf a} \bra{\sf c} + \text{e}^{2 \text{i} \Delta t } \ket{ \sf c} \bra{\sf
  d} \right) +  \text{H.c.},   
\end{align} 
for which the coupling $\ket{ \sf c} \bra{\sf d}$ is rapidly oscillating for
large $\Delta$.  This
enables the Hamiltonian to be approximated by 
\begin{align}
\hat{H}_{\text{eff}} = \frac{ \mathcal{E}_{\mu}}{\sqrt{2}} 
\ket{\sf a} \bra{\sf c}  + \text{H.c.},   
\end{align} 
which allows high-fidelity oscillations between states $\ket{ \sf a}$ and
  $\ket{\sf c}$.  The antisymmetric state can be prepared using
  $\mathcal{E}_{\mu,1} = -\mathcal{E}_{\mu,2} = \mathcal{E}_{\mu}$.  Note that
  this condition cannot be satisfied using a single laser field, because the
  distance between the atoms is much smaller than the wavelength
  $\lambda_0$.  One possibility is to use a standing-wave configuration, which
  is described in detail in Ref~\cite{Bei99}.  
\begin{figure}[t]
\begin{center}
\subfigure[]{\label{fig:pulses}
\includegraphics[width=6cm]{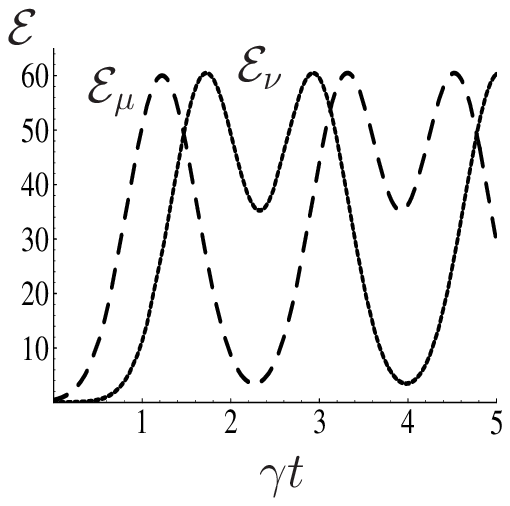}}
\subfigure[]{\label{fig:poprotstirap}
\includegraphics[width=6cm]{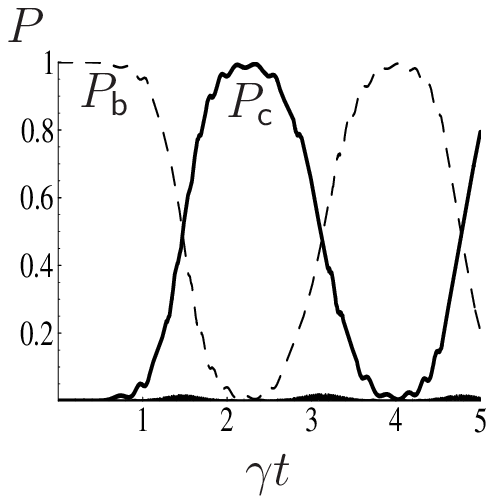}}
\end{center}
\caption{(a) Rabi frequencies of $\mathcal{E}_\mu = \sum_{n \in S}
\mathcal{E}$, where $\mathcal{E} =  60 
  \gamma \exp[-3.27 (t - n)^2]$ and $S = \{1.23, 3.31, 4.53, 6.62, \ldots  \}
  $, and $\mathcal{E}_\nu = \sum_{n \in S^\prime} \mathcal{E}$
for $S^\prime = \{1.72, 2.94,5.02,6.25,\ldots \}$ and 
  (b) populations of the levels {\sf a,b,c,d} when Rabi frequencies as
  depicted in (a) are applied to two atoms, initially in state $\ket{\sf b}$
  for $\alpha = 0$, $\xi = 1/5$, and $\omega_\delta = 0$.}         
\end{figure}

Once a single-excitation Bell-state is prepared, coherent Rabi oscillations
within the single-excitation subspace can be performed using bichromatic
fields. Here, we present two methods that admit this transformation, both of
which rely on neglecting rapidly oscillating terms.  We hope that 
these proposals will be of interest for proof-of-principle experiments
regarding coherent manipulation of atomic systems, particularly in the solid
state.  In particular, although the 2LAs are closely spaced, the
proposal here is comfortably within the capabilities of experiments performed
on nitrogen-vacancy centres in diamond~\cite{Me06}.   

We use two fields that satisfy $\mathcal{E}_{\mu,1} = \mathcal{E}_{\mu,2} =
\mathcal{E}_{\mu}$ and $\mathcal{E}_{\nu,1} = -\mathcal{E}_{\nu,2} =
\mathcal{E}_{\nu}$ respectively.  In Ref.~\cite{Bei99}, an example of an
experimental setup required to achieve these conditions is given. 
The field Hamiltonian in the collective basis can be written
\begin{align}
\label{eq:hamtot}
\hat{H}_{\text{eff}} =&  \frac{1}{\sqrt{2}} \Big(\mathcal{E}_\mu
  \text{e}^{-\text{i} t 
  \omega_\delta} \ket{\sf c} 
\bra{\sf d} 
+ \mathcal{E}_\nu \text{e}^{-\text{i} t \omega_\delta} \ket{\sf b} \bra{\sf d}
  + 
\mathcal{E}_\mu \text{e}^{-\text{i} t (2 \Delta + \omega_\delta)} \ket{\sf a}
  \bra{\sf c} - \mathcal{E}_\nu
\text{e}^{\text{i} t (2 \Delta - \omega_\delta)} \ket{\sf a} \bra{\sf b} \Big)
 +   \text{H.c.}, 
\end{align}
where $\omega_\mu = \omega_0 - \Delta - \omega_\delta$ and $\omega_\nu =
\omega_0 + \Delta - \omega_\delta$ for $\omega_\delta$ the detuning.  This
Hamiltonian is the same as that stated in Ref.~\cite{Bei99}.

Our first method applies the idea of single-atom two-photon Raman transitions
to the collective system of two atoms.  We show that coherent transfer between
Bell-states is made possible by judicious choice of detunings
and Rabi frequencies.  Figure~\ref{fig:ramantwo} shows the
effect of the bichromatic fields taking into account all four levels.  We
adiabatically eliminate the off-resonant levels to give 
\begin{align}
\hat{H}_{\text{eff}} = \mathcal{E}_\text{eff} \ket{\sf c} \bra{\sf b} + 
\text{H.c.},
\end{align}
for the effective Rabi
frequency 
\begin{align}
\mathcal{E}_\text{eff} = \frac{\mathcal{E}^*_\nu \mathcal{E}_\mu}{2
  \omega_\delta}, 
\end{align}
where we have neglected the fast oscillating terms that contribute to
$\mathcal{E}_\text{eff}$.  For the values quoted in Figure~\ref{fig:ramantwo},
$\mathcal{E}_\text{eff} = (15/100)\gamma$.
\begin{figure}[t]
\begin{center}
\subfigure[]{\label{fig:distf}
\includegraphics[width=6cm]{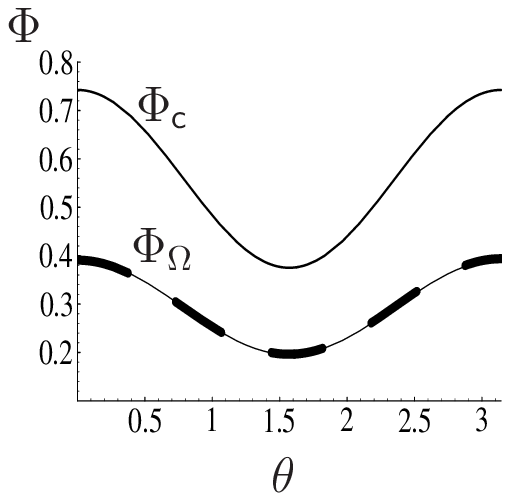}}
\subfigure[]{\label{fig:distfa}
\includegraphics[width=6cm]{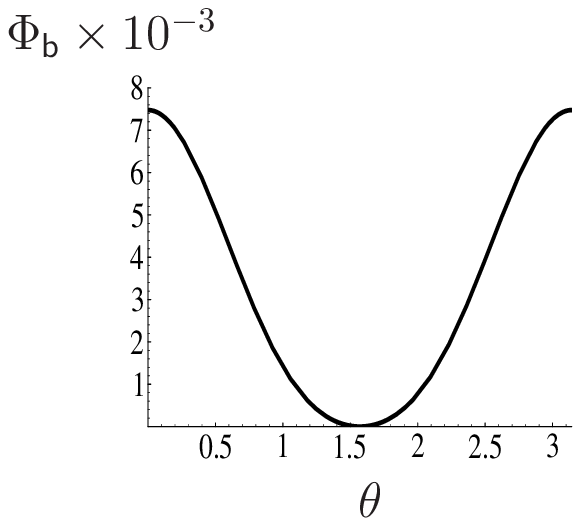}}
\end{center}
\caption{Average angular distribution of emitted photon with and without an
  incident field for $\alpha = \pi /2$, $\xi = 1/5$, and $0 <
  t < 20 
  \gamma^{-1}$, for emissions every $\gamma \delta t = 0.001$, and for
  $\ket{\psi(0)} = \ket{\sf c}$.  The
  distributions $\Phi_{\sf b}(\theta)$ and $\Phi_{\sf c}(\theta)$ are stated in
  Eqs.~\eqref{eq:pb} and 
  \eqref{eq:pc} respectively.  The angular distribution including the incident
  fields $\Phi_{\Omega}(\theta)$ is shown in Figure(a), for an exact
  numerical solution including all four levels (solid line) and for the ansatz 
 proposed in Eq.~\eqref{eq:ansatz} (dashed line) where $\Omega = \tanh
  t$. \label{fig:disttot} }         
\end{figure}

The second method for coherent transfer applies the ideas of
dark states in electromagnetically induced transparency (EIT)~\cite{Berg98} to
the collective states of two atoms.  This method of transfer is remarkably
robust, and allows  for high fidelity rotations at large interatomic
separations.  If the incident fields are set to resonance ($\omega_\delta =
0$) and the fast oscillating terms in Eq.~\eqref{eq:hamtot} are ignored, then
\begin{align} 
\label{eq:darkstate}
\ket{\psi} = \mathcal{E}_\nu \ket{\sf c} -\mathcal{E}_\mu \ket{\sf b} 
\end{align}
is an (unnormalised) eigenstate of $\hat{H}_{\text{eff}}$ that satisfies
$\hat{H}_{\text{eff}} \ket{\psi} = 0$.   The terms proportional to $\Delta$
can be ignored for approximately $\Delta \ge 2 \mathcal{E}_\nu, 2
\mathcal{E}_\mu$.  Similar terms were ignored (with numerical justification)
in Ref.~\cite{Bei99}.  For slowly-varying
fields, Eq.~\eqref{eq:darkstate} can be written  
\begin{align}
\ket{\psi (t)} = \cos \theta(t) \ket{\sf c} - \sin \theta(t) \ket{\sf b},
\end{align}
where $\tan \theta(t) = \mathcal{E}_\mu(t) /\mathcal{E}_\nu(t) $ and
slowly-varying means $ \dot{\theta} \ll \sqrt{\mathcal{E}^2_\mu(t) +
  \mathcal{E}^2_\nu (t)}$.  State $\ket{\psi (t)}$ allows for pulsed
population transfer between $\ket{\sf c}$ and $\ket{\sf b}$.
We solve the full Hamiltonian numerically, and find
that $\ket{\psi (t)}$ is a good approximation to an eigenstate of
$\hat{H}_{\text{eff}}$.  The population of levels $\ket{\sf b}$ and $\ket{\sf
  c}$ for the canonical counter-intuitive pulse sequence is shown in 
Figure~\ref{fig:stiraptwo}.  As noted in the caption, the transfer was possible
at separations $\simeq (1/5) \lambda_0$.  This
regime is accessible to present experiment~\cite{Me06}.    Note the
time taken for rotation is large, because when $\xi = 1.25$,  $\Delta
\simeq 1.2 \gamma$.  

In order to coherently transfer population between the Bell-states, the
Rabi frequencies of the incident fields are modelled as sums of exponentials
[Figure~\ref{fig:pulses}].  The population of all four levels is shown in
Figure~\ref{fig:poprotstirap}, which shows that high-fidelity transfer is
possible using time-varying incident fields.  The time taken for rotation is
comparable with the emission time from a 2LA in free space.  
%%%%%%%%%%%%%%%%%%%%%%%%%%%%%%%%%%%%%%%%%%%%%%%%%%%%%%%%%%%%%%%%%%%%%%%%%%%%%
\subsection{Emission properties}
\label{sec:emit2a}
%%%%%%%%%%%%%%%%%%%%%%%%%%%%%%%%%%%%%%%%%%%%%%%%%%%%%%%%%%%%%%%%%%%%%%%%%%%%
We are interested in the effect of incident fields on the collective
spontaneous-emission time and pattern from two atoms.  We focus on 
the emission pattern from the single-excitation subspace, whose state is given
by
\begin{align}
\ket{\bar{\psi}(t)} = \text{e}^{-\text{i} \hat{H}_{\text{eff}} t}
\ket{\psi(0)}, 
\end{align}
where $\hat{H}_{\text{eff}} = \hat{H}_S + \hat{H}_I$ , $\ket{\psi(0)}$ is
a linear combination of $\ket{\sf b}$ and $\ket{\sf c}$, and
$\ket{\bar{\psi}}$ is unnormalised. 
The angular distribution of the emitted photon is~\cite{Car00,Clem03} 
\begin{align}
\Phi(\theta) \sin \theta d \theta = \int_0^\infty \text{d} t
\bra{\bar{\psi}(t)}\hat{S}^\dagger(\theta) \hat{S}(\theta)
\ket{\bar{\psi}(t)}.  
\end{align}
We denote the angular distribution affected by an incident field with 
$\Phi_\Omega(\theta)$.  Without an incident field, the angular distribution of
emission from states 
$\ket{\sf b}$ and $\ket{\sf c}$ can be written~\cite{Clem03} 
\begin{align}
\Phi_{\sf b}(\theta) &= D(\theta) \sin ( \pi s \cos \theta)^2, \label{eq:pb}
\\ 
\Phi_{\sf c}(\theta) &= D(\theta) \cos ( \pi s \cos \theta)^2. 
\label{eq:pc}
\end{align}
Figure~\ref{fig:disttot} shows the angular distribution of emission from
$\ket{\sf b}$ and $\ket{\sf c}$ for an interatomic separation $\xi =
1/5$.  We have normalised the distributions so that $\Phi_{\sf
  b}(\theta) +  \Phi_{\sf c}(\theta) = D(\theta)$.  

In light of the manipulation methods proposed in Sec.~\ref{sec:manip2a}, we
propose the ansatz  
\begin{align}
\label{eq:ansatz}
\hat{H}_{\text{I}} = \Omega \ket{\sf c}\bra{ \sf b} + \text{H.c.},
\end{align} 
where $\Omega$ is the Rabi frequency.  Figure~\ref{fig:disttot} compares the
angular distribution when using the ansatz to the exact numerical solution.
For the parameters quoted in Figure~\ref{fig:disttot}, the ansatz is a good
approximation.  The Rabi frequency $\Omega = \tanh t$ in order to 
account for the dark state preparation time shown explicitly for small $t$ in
Figure~\ref{fig:poprotstirap}.  

The angular distribution with an incident field is not the dipole emission
pattern.  The field causes the emission patterns from $\ket{\sf b}$ and
$\ket{\sf c}$ to mix.  Using the ansatz and 
\begin{align}
\ket{\psi(t)} = \cos \Omega t\ket{\sf c} - \text{i} \sin \Omega t\ket{\sf b},
\end{align}
the emission pattern for $\ket{\psi(0)} = \ket{\sf c}$ is
\begin{align}
\Phi_\Omega(\theta) \sin \theta d \theta =& \int_0^\infty \text{d} t \frac{
    D(\theta) 
    (\gamma + \Gamma)  \text{e}^{- \gamma t}}{\Gamma^2 - 4 \Omega^2} \big[
    \Omega_R 
    \cos \frac{\zeta}{2} \cosh \frac{\Omega_R t}{2} - (\Gamma
    \cos\frac{\zeta}{2} + 2 \Omega \sin \frac{\zeta}{2})\sinh 
     \frac{\Omega_R t}{2} \big]^2,
\end{align} 
for $\Omega_R \equiv \sqrt{\Gamma^2 - 4 \Omega^2}$ and $\zeta \equiv 2 \pi s
\cos \theta$, and where we have normalised so that when $\Omega \to 0$,
Eq.~\eqref{eq:pc} is recovered.  The matrix element
\begin{align}
\label{eq:matel}
\bra{\psi(t)}\hat{S}^\dagger(\theta) \hat{S}(\theta)
\ket{\psi(t)} = 2 \gamma D(\theta) \cos (\Omega  t + \pi s \cos  \theta)^2,
\end{align}
for $\ket{\psi(0)} = \ket{\sf c}$, explicitly shows the dependence of
emission direction on $\Omega$.  The average of
Eq.~\eqref{eq:matel} for all times gives the (unnormalised) angular 
distribution of emission. Note that the emission is not simply $\Phi_{\sf b}
|c_{\sf b}(t)|^2 + \Phi_{\sf c}|c_{\sf c}(t)|^2$ for $\ket{\psi(t)} = c_{\sf
  b}(t) \ket{\sf b} + c_{\sf c}(t) \ket{\sf c}$.   

The emission time is also altered by the presence of an incident field.
Explicitly including decay, for $\Omega \gg \Gamma$, 
\begin{align}
\ket{\psi(t)} =& \text{e}^{-t \gamma /2}(c_{\sf c} \cos \Omega t - \text{i}
c_{\sf b} \sin 
\Omega t ) \ket{\sf c} + \text{e}^{-t \gamma /2}(c_{\sf b} \cos \Omega t -
\text{i} c_{\sf  c} \sin  
\Omega t ) \ket{\sf b},
\end{align}       
where $\ket{\psi(0)} = c_{\sf b} \ket{\sf b} + c_{\sf c} \ket{\sf c}$.  This
shows that for large $\Omega$, the decay rate of two atoms becomes equal to
that expected from a single atom.  
\begin{figure}[tp]
\begin{center}
\includegraphics[width=6cm]{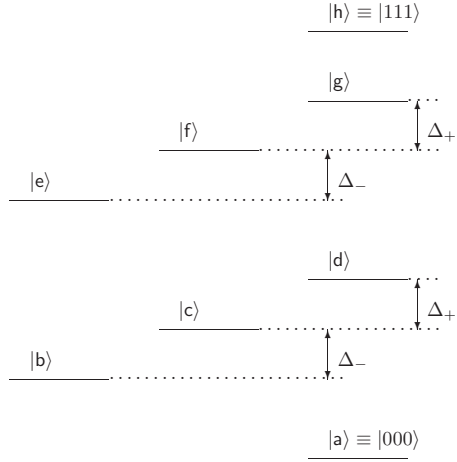}
\end{center}
\caption{\label{fig:dipsplit}Energy level scheme of three atoms, with $
  \Delta_{\pm} = (1 / 2)(\delta \pm 3\Delta_{13})$, for $\delta \equiv
  \sqrt{8 \Delta_{12}^2  +  \Delta_{13}^2}$.  In 
  the Dicke limit~\cite{Dicke}, $\ket{\sf b} = (1/\sqrt{6})
  (-2\ket{001} +  \ket{010} + \ket{100})$ and $\ket{\sf c} =
  (1/\sqrt{2}) (\ket{010} - \ket{100}) $. }     
\end{figure}

We have shown that for two atoms the emission direction and time is altered by
an incident field that rotates within the one-excitation subspace.
Due to the speed and fidelity of the rotation {\sf b-c} using
pulsed-population-transfer, and the large separations at which it is possible,
we hope these effects are observable with present experiments.
%%%%%%%%%%%%%%%%%%%%%%%%%%%%%%%%%%%%%%%%%%%%%%%%%%%%%%%%%%%%%%%%%%%%%%%%%%%%%
\section{Three atoms}
\label{sec:that}
%%%%%%%%%%%%%%%%%%%%%%%%%%%%%%%%%%%%%%%%%%%%%%%%%%%%%%%%%%%%%%%%%%%%%%%%%%%%
The spontaneous emission from two atoms in the presence of coherent rotations
allows some insight into the effect of incident fields on emission
characteristics.  However, two atoms is too few for both observation of
directional superradiance and for exploiting collective decay properties
for quantum information processing. So, we extend the analysis to three atoms.

The three atoms are arranged in a linear configuration according to
$\boldsymbol{r}_{1}= 0$, $\boldsymbol{r}_{2}= \boldsymbol{r}$ 
and $\boldsymbol{r}_{3}= 2\boldsymbol{r}$.  The energy level scheme is shown in
Figure~\ref{fig:dipsplit}.  Note that the definitions of $\ket{\sf a}, \ket{\sf
  b}, \ket{\sf c}$, and $\ket{\sf d}$ have altered from Sec.~\ref{sec:tat}.
The symmetric states are labelled $\ket{\sf d}$ and
$\ket{\sf g}$.  For quantum information processing, a qubit is most sensibly
encoded using $\{ \ket{\sf b}, \ket{\sf c} \}$ as logical states.  These are
the slowest decaying excited states in the eight-level system, and in the
limit $\xi \to 0$ have infinite decay times.  They form the lower states of a
decoherence-free subsystem~\cite{Kem01}.
%%%%%%%%%%%%%%%%%%%%%%%%%%%%%%%%%%%%%%%%%%%%%%%%%%%%%%%%%%%%%%%%%%%%%%%%%%%%%
\subsection{Manipulation}
\label{sec:manip3a}
%%%%%%%%%%%%%%%%%%%%%%%%%%%%%%%%%%%%%%%%%%%%%%%%%%%%%%%%%%%%%%%%%%%%%%%%%%%%
Recently, there has been a proposal for the preparation, rotation, and
readout of a qubit encoded across $\{ \ket{\sf b}, \ket{\sf c}
\}$~\cite{Brooke07}.  So, we assume that $\ket{\sf b}$ can be prepared to high
fidelity, with the effective coupling in the interaction picture with respect
to the Hermitian part of $\hat{H}_{\text{S}}$ given by
\begin{align}
\mathcal{E}_{\text{eff}} =& \text{e}^{-\text{i} \boldsymbol{k}_\mu \cdot
  \boldsymbol{r}}  
  \sqrt{1 - \frac{\Delta_{13}}{\delta}} \frac{\mathcal{E}_\mu}{2 \kappa} (3
  \Delta_{12} 
  \Delta_{13} -  \delta \Delta_{12} + 2 \kappa \cos \boldsymbol{k}_\mu \cdot
  \boldsymbol{r}),  
\end{align}
for $\mathcal{E}_\mu = \boldsymbol{d} \cdot \boldsymbol{E}_\mu$, $\delta
  \equiv \sqrt{8 
  \Delta_{12}^2  +  \Delta_{13}^2}$, and $\kappa \equiv 2 \Delta_{12}^2 +
  \Delta_{13}(\Delta_{13} - \delta)$.  The laser frequency is $\omega_\mu =
  \omega_{\sf ab} =  (1/2) (\Delta_{13} -   \delta) + \omega_0$.
Population transfer can be performed by using resonant fields and
counter-intuitive pulse sequences.  For this method, the incident fields
resonantly couple {\sf b-g} and {\sf c-g}.  So, by choosing the 
laser frequencies $\omega_\nu = \omega_{\sf bg} = \delta + \omega_0$ and
$\omega_\mu = \omega_{\sf cg} = (1/2)(3 
\Delta_{13} + \delta) + \omega_0$ 
in the interaction picture with respect to the Hermitian part of
$\hat{H}_{\text{S}}$ the coupling {\sf c-g} is
\begin{align}
\mathcal{E}_{\sf cg} =& \frac{ \Delta_{12} (\delta + 3 \Delta_{13})
  \sqrt{\delta + \Delta_{13}}}{2 \sqrt{2 \delta} [ 2 \Delta_{12}^2 +
  \Delta_{13} (\delta + \Delta_{13}) ]}
\big[ \text{e}^{-\text{i}(4 \boldsymbol{k} \cdot
  \boldsymbol{r} + 3 
  \Delta_{13} t )/2} (\text{e}^{2 i \boldsymbol{k} \cdot \boldsymbol{r}} - 1)(
  \text{e}^{ i 
  \delta t/2} 
  \mathcal{E}_\nu + \text{e}^{3 i \Delta_{13} t /2} \mathcal{E}_\mu ) \big],
\end{align}
for $\boldsymbol{k}_\nu = \boldsymbol{k}_\mu = \boldsymbol{k}$, and the
coupling {\sf b-g} 
is  
\begin{figure}[t]
\begin{center}
\subfigure[]{\label{fig:pulse3}
\includegraphics[width=6cm]{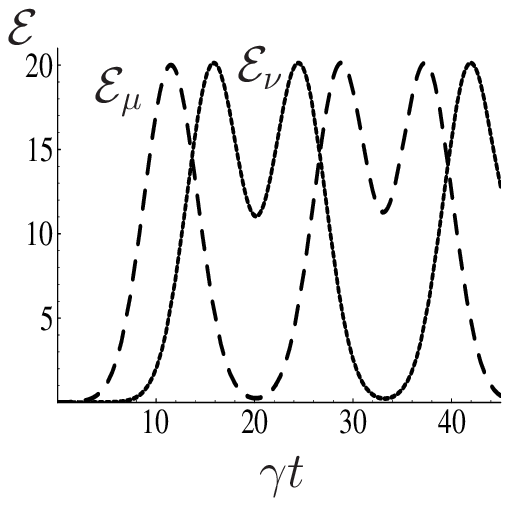}}
\subfigure[]{\label{fig:stirap3}
\includegraphics[width=6cm]{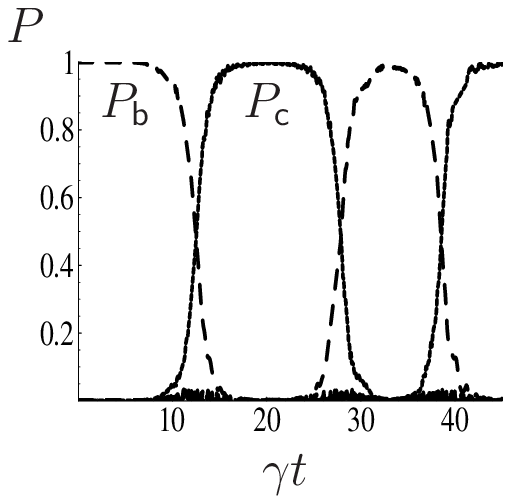}}
\end{center}
\caption{ (a) Rabi frequencies of $\mathcal{E}_\mu = \sum_{n=1,2.5,3.25,4.8,
    \ldots } 20 \gamma \exp[-(9/100)(t - n t_\mu)^2/T^2]$ and
    $\mathcal{E}_\nu = \sum_{n=1,1.55,2.65, \ldots } 20 \gamma
    \exp[-(9/100)(t - n t_\nu)^2/T^2]$ for $T = 1.15 \gamma^{-1}$,
    $t_\mu = 10 T$, $t_\nu = t_\mu + 3.75T$, and laser frequencies
    of $\omega_\nu = \omega_{\sf bg} = \delta + \omega_0$ and $\omega_\mu =
    \omega_{\sf cg} = (1/2)(3 \Delta_{13} +  
    \delta ) + \omega_0$, and (b) populations of levels {\sf a, \ldots, h}
    when pulses of 
    type (a) are applied to the collective eight-level system that is
    initially in $\ket{\sf b}$ for $\alpha = \pi/2$ and the
    separation between neighbouring atoms is $\xi =
    1/5$. \label{fig:str3at} 
}
\end{figure}
\begin{align}
\mathcal{E}_{\sf bg} =& \frac{\sqrt{\delta^2 - \Delta_{13}^2}
    \text{e}^{-\text{i} 
    \boldsymbol{k} 
    \cdot \boldsymbol{r}}(  \mathcal{E}_\nu +
    \text{e}^{-\text{i}(\delta - 3 \Delta_{13})t/2} 
\mathcal{E}_\mu)}{2  \delta \kappa [ 2 \Delta_{12}^2 +
  \Delta_{13} (\delta + \Delta_{13}) ]} \big[ (2 \Delta_{12}^2 + \Delta_{13}^2
  )^2 
 - \delta^2 \Delta_{13}^2 + \varsigma \cos \boldsymbol{k} \cdot \boldsymbol{r}
    \big], 
\end{align}
for $\kappa \equiv 2 \Delta_{12}^2 + \Delta_{13}(\Delta_{13} - \delta)$, and
$\varsigma \equiv 2 \Delta_{12} \Delta_{13} ( 6 \Delta_{12}^2 - \delta^2 + 3
\Delta_{13}^2)$.  These couplings are such that, with judicious choice of Rabi
frequency, they permit coherent Rabi oscillations between {\sf b-g} and {\sf
  c-g}.  Setting all other couplings to zero, the dark state is written 
\begin{align}
\ket{\psi} = \mathcal{E}_{\sf cg} \ket{\sf b} - \mathcal{E}_{\sf bg} \ket{\sf
  c}.
\end{align}
As for the two-atom case, this is not an eigenstate of the total field
Hamiltonian. Figure~\ref{fig:stirap3} shows the population of all eight levels
when the pulse sequence shown in Figure~\ref{fig:pulse3} is applied to the
collective system. Note that for three atoms, the laser fields are
co-propagating so $\boldsymbol{r} \parallel \boldsymbol{k}$.  Surprisingly,
even in the presence of all the other couplings, there is a regime that admits
coherent population transfer using counter-intuitive bichromatic fields.  

Similar to Sec.~\ref{sec:manip2a}, the manipulation can also be performed  
using collective two-photon Raman transitions~\cite{Brooke07}.  Both methods
of rotation rely on the splitting due to the dipole-dipole 
interaction being large enough to enable incident fields to isolate the
collective transitions.  So, for experimental implementations, closely-spaced
2LAs are required.  
%%%%%%%%%%%%%%%%%%%%%%%%%%%%%%%%%%%%%%%%%%%%%%%%%%%%%%%%%%%%%%%%%%%%%%%%%%%%%
\subsection{Emission properties}
\label{sec:emit3a}
%%%%%%%%%%%%%%%%%%%%%%%%%%%%%%%%%%%%%%%%%%%%%%%%%%%%%%%%%%%%%%%%%%%%%%%%%%%%
\begin{figure}[b]
\begin{center}
\includegraphics[width=6cm]{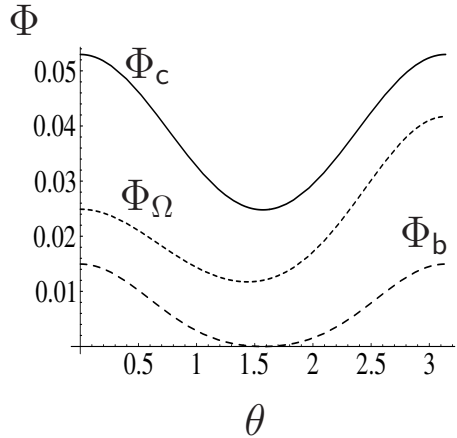}
\end{center}
\caption{Average angular distribution of emitted photon with and without an
  incident field for $\alpha = \pi/2$, $\xi = 1/5$, and $0 <
  t < 70 
  \gamma^{-1}$, for emissions every $\gamma \delta t = 0.01$, and for
  $\ket{\psi(0)} = \ket{\sf b}$, normalised to the emission
  rate. \label{fig:3dr} }         
\end{figure}
Using the ansatz  
\begin{align}
\label{eq:ansatz2}
\hat{H}_{\text{I}} = \Omega \ket{\sf c}\bra{ \sf b} + \text{H.c.},
\end{align}
where $\Omega$ is the Rabi frequency, the decay of
the levels $\ket{\sf b}$ and $\ket{\sf c}$ for
$\Omega \gg \gamma_{\sf b}, \gamma_{\sf c}$ is $\exp[-(1/2) (\gamma_{\sf
    b} + \gamma_{\sf c}) t]$, where $\gamma_{\sf b}$ and $\gamma_{\sf c}$ are
the decay widths of $\ket{\sf b}$ and $\ket{\sf c}$ respectively.  

Comparing $\hat{S}(\theta) \ket{\psi(t)}$, where $\ket{\psi(t)} = \exp[
  -\text{i} 
  \hat{H}_{\text{eff}} t] \ket{\psi(0)}$ for
$\hat{H}_{\text{eff}} = \hat{H}_I$ from Eq.~\eqref{eq:ansatz2} and
  $\ket{\psi(0)} = \ket{\sf b}$, with $\hat{S}(\theta) \ket{\sf b}$ 
allows us to quantify the effect of the rotation on the emission direction. So,  
\begin{align}
\hat{S}(\theta) \ket{\sf b} =& \text{e}^{-2 \text{i} \zeta}
  \frac{\sqrt{D(\theta) 
  \gamma}}{2 \kappa} \sqrt{1 - \frac{\Delta_{13}}{\delta}}(\kappa -
  \text{e}^{\text{i} \zeta} \eta) \ket{\sf a},  
\end{align}
where $\eta \equiv (\Delta_{12} +
\Delta_{13})(\delta - 2 \Delta_{12} - \Delta_{13})$, and
\begin{align}
\hat{S}(\theta)\ket{\psi(t)} =& \text{e}^{-2 \text{i} \zeta}
  \frac{\sqrt{D(\theta) 
  \gamma}}{2 \kappa} \bigg( \sqrt{1 - \frac{\Delta_{13}}{\delta}} (\kappa -
  \text{e}^{\text{i}  \zeta} \eta) \cos \Omega t +  \text{i} \sqrt{2}
  (\text{e}^{\text{i}  \zeta} -1)\kappa \sin \Omega t 
  \bigg) \ket{\sf a}. 
\end{align}
The emission pattern has been altered by the field Hamiltonian, and 
includes a contribution from
\begin{align}
\hat{S}(\theta) \ket{\sf c} = \text{e}^{-2 \text{i}  \zeta}(1 - \text{e}^{
  \text{i}  \zeta}) 
\sqrt{\frac{D(\theta) \gamma}{2} } \ket{\sf a},
\end{align}
dependent on the time of emission and the size of $\Omega$.
Figure~\ref{fig:3dr} shows the numerical solution for the emission direction
for 
the incident fields shown in Figure~\ref{fig:str3at}.  The directional emission
is proportional to the populations in $\ket{\sf b}$ and $\ket{\sf c}$.  So,
the directional emission is affected by the incident fields similarly to two
2LAs, and thus serves as a measure of the efficacy of population transfer.
%%%%%%%%%%%%%%%%%%%%%%%%%%%%%%%%%%%%%%%%%%%%%%%%%%%%%%%%%%%%%%%%%%%%%%%%%%%%%
\section{Conclusion}
\label{sec:conc}
%%%%%%%%%%%%%%%%%%%%%%%%%%%%%%%%%%%%%%%%%%%%%%%%%%%%%%%%%%%%%%%%%%%%%%%%%%%%
We have applied the ideas of STIRAP to multi-atom systems, and found that
counter-intuitive pulse sequences can be used to transfer population,
even if there is no null eigenstate of the field Hamiltonian.  For two
2LAs, this method of transfer is robust at large separations, and so
particularly amenable to experiment.

For two and three 2LAs, we showed how the direction and time of the
spontaneous emission is altered in the presence of an incident field.  We gave
analytical expressions for the directional matrix elements 
that depend on the strength of the incident fields. For three 2LAs we showed
how the directional emission  can give an indication of the efficacy of a
rotation within a decoherence-free subsystem.  We hope the results
presented here will be useful for proof-of-principle
experiments involving small numbers of dipole-coupled 2LAs. 
%%%%%%%%%%%%%%%%%%%%%%%%%%%%%%%%%%%%%%%%%%%%%%%%%%%%%%%%%%%%%%%%%%%%%%%%%%%%% 
\section*{Acknowledgements}
%%%%%%%%%%%%%%%%%%%%%%%%%%%%%%%%%%%%%%%%%%%%%%%%%%%%%%%%%%%%%%%%%%%%%%%%%%%%%
We especially thank James Clemens for many helpful emails. We
also thank Jim Cresser, Karl-Peter Marzlin, and Barry Sanders for
helpful discussions, and Stojan Rebic for comments on the manuscript.  This
work was supported by Macquarie University. 
\bibliography{ref}
\end{document}